# Structural and magnetic properties of a new and ordered quaternary alloy MnNiCuSb (SG: $F\bar{4}3m$)

Zeba Haque[1], Gohil S Thakur[1], Somnath Ghara[2], L C Gupta[1,†], A Sundaresan[2] and A K Ganguli[1,3]*

[1] Department of Chemistry, Indian Institute of Technology, New Delhi, India, 110016

[2] Chemistry and Physics of Materials Unit, JNCASR, Jakkur, Bangalore, India,

[3] Institute of Nano Science & Technology, Habitat Centre, Punjab, India, 160062

*Email: ashok@chemistry.iitd.ac.in

**Abstract**

We have synthesized a new crystallographically ordered quaternary Heusler alloy, MnNiCuSb. The crystal structure of the alloy has been determined by Rietveld refinement of the powder x-ray diffraction data. This alloy crystallizes in the LiMgPdSb type structure with $F\bar{4}3m$ space group. MnNiCuSb is a ferromagnet with a high $T_C$ ~ 690K and magnetic moment of 3.85µB/f.u. Besides this we have also studied two other off-stoichiometric compositions; one Cu rich and the other Ni rich ($MnNi_{0.9}Cu_{1.1}Sb$ and $MnNi_{1.1}Cu_{0.9}Sb$) which are also ferromagnets. It must be stressed that MnNiCuSb is one of the very few known, non-Fe containing quaternary Heusler alloys with 1: 1: 1: 1 composition.

## Introduction

Since their discovery, Heusler alloys have been a fascinating area of research for both experimental and theoretical studies because of their interesting magnetic properties. The remarkable property of conventional Heusler alloys, such as $Cu_2MnAl$, is that none of their constituent elements is ferromagnetic but the alloys themselves are ferromagnetic with a high $T_C$ [1]. There is another class of regular Heusler alloys which are quaternary alloys and belong to the LiMgPdSb type structure (space group $F\bar{4}3m$) [2]. Alloys such as NiMnSb have been known in literature as half Heusler alloys. They are structurally very similar to the conventional Heusler alloys. An outstanding feature of these alloys, as pointed out by Groot et al, is that they are half-metallic ferromagnets (HMFs) possessing 100% spin polarization at the Fermi level [3]. They are characterized by coexistence of metallic electronic structure for electron spin in one direction and semiconducting for the electron spin in the opposite direction. The magnetic properties are sensitive to chemical order, composition of the sample [4, 5] and the heat treatment [6]. Later on, many regular Heusler alloys such as $Co_2CrAl$, $Mn_2CuSb$, $Ti_2FeAl$, $Ti_2CoGa$, $Co_2MnSi$, $Co_2MnGe$ etc. have also been predicted to be half metallic ferromagnet [7-9]. Pseudo-ternary Heusler alloys such as $Co_2Cr_{1-x}Fe_xAl$ [10], $Co_2Mn_{1-x}Fe_xSi$ [11] have also been studied for their varied chemical properties.

Recently, a number of ternary half-Heusler compounds have been predicted to be three-dimensional (3D) topological insulator candidates due to their topologically nontrivial band inversion [12-14]. A combination of magnetic Heusler alloy materials (for example $Co_2FeSi$ thin films deposited on metallic substrate) opens up a new area of spintronics [15, 16]. Polycrystalline Heusler alloy $Co_2FeAl$ fabricated on $Si/SiO_2$ amorphous substrate is known to possess high tunnel magnetoresistance ratio [17].

The regular Heusler alloy of the type $X_2YZ$, crystallizes in cubic $Cu_2MnAl$ type structure ($L2_1$ structure) with space group $Fm\bar{3}m$ [3, 18]. Here X and Y are transition metals and Z is an sp-element. It consists of four interpenetrating fcc sublattices. X atoms occupy 4a (0, 0, 0) and 4b (0.5, 0.5, 0.5) positions; Y atoms occupy the 4c (0.25, 0.25, 0.25) and Z atoms 4d (0.75, 0.75, 0.75) positions. It is also possible that Z and Y atoms occupy the 4a and 4c sites respectively, while the two X atoms reside at 4b and 4d sites. An ordered structure is obtained due to the non-mixing of atoms at these sites. The half Heusler alloy (XYZ type) crystallizes in cubic NiMnSb type structure ($C1_b$ structure) with space group $F\bar{4}3m$. One of the X sites is vacant in the half Heusler alloys. Thus the $C1_b$ structure, can be derived from $L2_1$ structure by removing half of the X sites in an ordered manner. However, in the structure, atomic disorder can be introduced by partial substitution of X atoms. For example, partial substitution of Cu by Ni at 4a site has been studied in $Cu_{1-x}Ni_xMnSb$ [19].

Another class of Heusler alloys which are quaternary 1:1:1:1 type which belong to the LiMgPdSb type structure (space group $F\bar{4}3m$) [2]. In such Heusler alloys, one of the X atoms is completely substituted by another transition metal element. There are only few examples known for such quaternary alloys, CoFeMnZ (Z = Al, Ga, Si and Ge) [20, 21], CoFeCrAl [22], CoFeMnSi [23] NiCoMnZ (Z = Al, Ge and Sn) [24] NiFeMnGa, NiCoMnGa and CuCoMnGa [25] $CoFe_{1+x}Ti_{1-x}Al$ and $CoMn_{1+x}V_{1-x}Al$ [26, 27]. Ordered quaternary Heusler alloys are formed when the occupancy of each atom at thefour Wyckoff positions is unity.

## Experimental

In this paper we are presenting new quaternary Heusler alloy MnNiCuSb. We have studied three compositions $MnNi_{1-x}Cu_{1+x}Sb$ (x = 0.1, 0 and -0.1). The alloys $MnNi_{1-x}Cu_{1+x}Sb$ (x = 0.1, 0 and -0.1) were synthesized via two step solid state method. Stoichiometric amount of Mn, Ni, Cu and Sb metal powders was weighed, ground and pelletized inside an argon filled glove box ($H_2O$, $O_2$ < 0.1 ppm). The pellet was sealed in an evacuated quartz tube under high vacuum (pressure ~ $10^{-4}$ torr) and was sintered at 1000°C for 24 hours. To ensure phase homogeneity, the samples were sintered again at 700°C for 24 hours followed by quenching in water. The phase purity was checked by powder X-ray diffraction technique (PXRD) (Bruker D8 Advance) using Cu-Kα radiation. Structural refinement of the powder X-ray diffraction data was carried out by Rietveld method using TOPAZ package [28]. Magnetic measurements were carried out on a SQUID vibrating magnetometer (VSM) and a high temperature VSM in Physical Property Measuring System (PPMS), Quantum Design, USA. Microstructrural studies of MnNiCuSb was analyzed with the aid of field-emission scanning electron microscopy (FESEM) (FEI QUANTA 3D FEG) operating at an accelerating voltage of 5 kV equipped with an energy dispersive X-ray spectroscopy (EDS) detector for elemental mapping.

## Results and Discussion

The powder X-ray diffraction pattern of

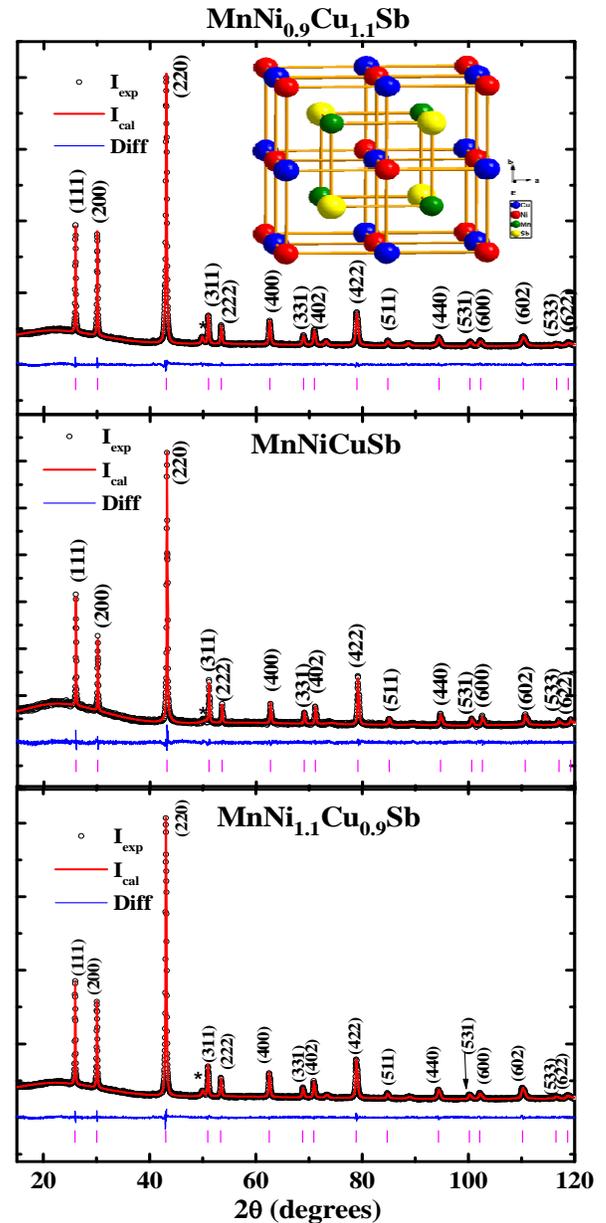

Figure 1. Rietveld refinement of powder X-ray diffraction pattern of $MnNi_{1-x}Cu_{1+x}Sb$ (x = 0.1, 0 - 0.1). Inset shows the crystal structure of MnNiCuSb. Yellow, green, red and blue balls correspond to Sb, Mn, Ni and Cu atoms respectively.

$MnNi_{0.9}Cu_{1.1}Sb$, MnNiCuSb and $MnNi_{1.1}Cu_{0.9}Sb$ collected at room temperature shows that all the three compounds crystallize in

the cubic Heusler structure. The XRD pattern could be indexed to cubic structure with space group $F\bar{4}3m$. Rietveld structural refinement of the PXRD is shown in figure 1. The crystallographic details of the refinement are listed in Table I. Small impurity peaks (<5%) corresponding to Cu metal are present in all the compositions.

| Crystallographic details of $MnNi_{0.9}Cu_{1.1}Sb$ | | | | | | |
|---|---|---|---|---|---|---|
| a (Å) = 5.93667(6); $R_{wp}$ = 2.36%; $\chi^2$ = 1.37 | | | | | | |
| Site | x | y | z | Atom | Occ | Beq |
| 4a | 0 | 0 | 0 | Ni/Cu | 0.90 (2)/0.10(2) | 1.50 (2) |
| 4c | 0.25 | 0.25 | 0.25 | Cu | 1.00 (6) | 1.50 (6) |
| 4b | 0.5 | 0.5 | 0.5 | Mn | 1.00 (2) | 1.50 (1) |
| 4d | 0.75 | 0.75 | 0.75 | Sb | 0.95 (3) | 1.20 (3) |
| Crystallographic details of MnNiCuSb | | | | | | |
| a (Å) = 5.92382(9); $R_{wp}$ = 4.01%; $\chi^2$ = 1.27% | | | | | | |
| Site | x | y | z | Atom | Occ | Beq |
| 4a | 0 | 0 | 0 | Ni | 1.002 (4) | 1.02 (4) |
| 4c | 0.25 | 0.25 | 0.25 | Cu | 0.962 (5) | 1.01 (2) |
| 4b | 0.5 | 0.5 | 0.5 | Mn | 1.009 (5) | 1.01 (4) |
| 4d | 0.75 | 0.75 | 0.75 | Sb | 1.038 (4) | 0.90 (1) |
| Crystallographic details of $MnNi_{1.1}Cu_{0.9}Sb$ | | | | | | |
| a (Å) = 5.94063(8); $R_{wp}$ = 2.48%; $\chi^2$ = 1.43 | | | | | | |
| Site | x | y | z | Atom | Occ | Beq |
| 4a | 0 | 0 | 0 | Ni | 1.000 (8) | 1.50 (2) |
| 4c | 0.25 | 0.25 | 0.25 | Cu/ Ni | 0.90 (1)/ 0.10 (1) | 1.50 (8) |
| 4b | 0.5 | 0.5 | 0.5 | Mn | 1.00(2) | 1.50 (1) |
| 4d | 0.75 | 0.75 | 0.75 | Sb | 0.950(1) | 1.20 (4) |

Table 1. Crystallographic details of $MnNi_{1-x}Cu_{1+x}Sb$. MnNiCuSb has the smallest lattice parameter as compared to those of the Cu-rich and Ni-rich phases. The ionic radius of $Ni^{2+}$ ion (0.69Å) is slightly smaller than that of $Cu^{1+}$ ion (0.77Å) in the same coordination number. Thus we obtain the expected trend. We fixed the position of Mn and Sb atoms at 4b and 4d sites respectively. Ni and Cu atoms occupy 4a and 4c sites. This atomic arrangement yields good reliability factors ($R_{wp}$ and $\chi^2$) in all the three cases. Identification of ordered disordered structure by powder X-ray diffraction is carried out by the relation between atomic ordering and the variation of intensities of the diffraction lines [29]. The superlattice reflections (1, 1, 1) and (2, 0, 0) with moderate intensity are clearly visible in PXRD, which hints towards ordered atomic arrangement. However it is difficult to confirm the ordering in structure based on the X-ray diffraction studies as the diffraction from Cu, Ni and Mn atoms is identical due to similar scattering factor. Neutron diffraction studies would give correct ordering in these structures.

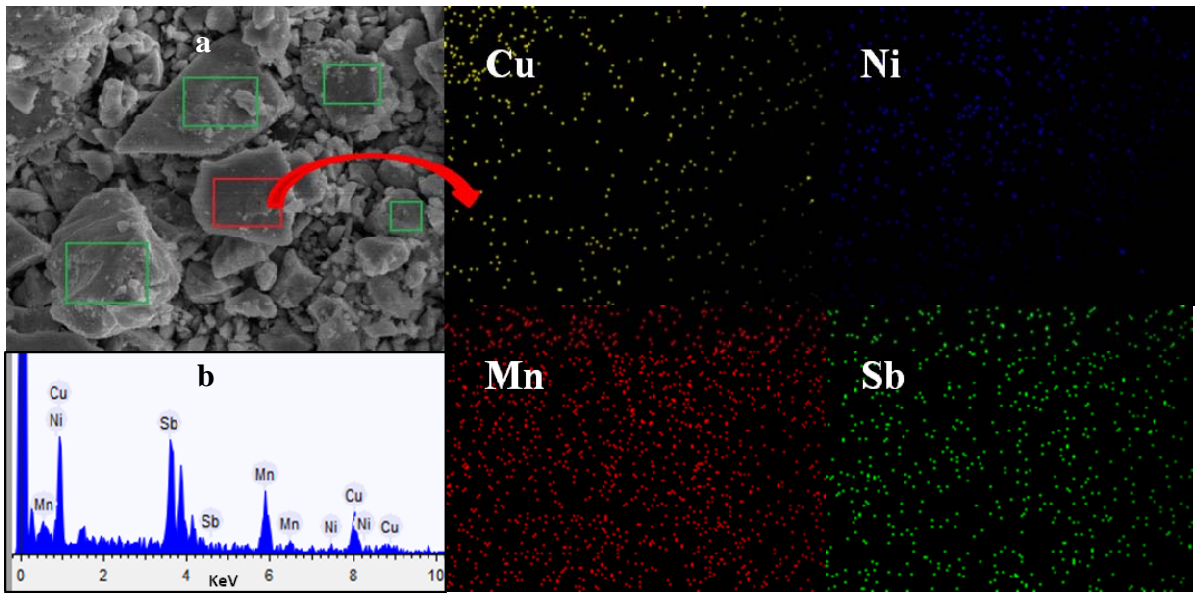

Figure 2 (a) Magnified FESEM image of various grains and mapping result of the rectangular area in red confirms presence of all four elements. (b) FESEM EDS analysis of MnNiCuSb showing the presence of Mn, Ni, Cu and Sb.

Elemental composition, the amount of distribution of elements present in MnNiCuSb and any phase separation at microscopic scale was confirmed by energy dispersive X-ray spectroscopy (EDS) elemental mapping was conducted (Fig. 2). The final stoichiometry was calculated by averaging the data collected over many grains (green and red rectangular areas) was found to be $Mn_{0.97}NiCu_{0.95}Sb$ which is close to the nominal composition MnNiCuSb.

The mapping results of the selected rectangular area in red color of FESEM image of a single grain of MnNiCuSb clearly depicts the coexistence of Mn, Ni, Cu and Sb elements in the alloy. We can see that the distribution of all the elements is homogeneous and uniform. No other binary or ternary phases could be detected. We also tried other compositions by varying the pnictide atom viz. MnNiCuPn (Pn = P, As and Bi) but none of these form. So along with the charge the size of Sb atom stabilizes the formation of cubic structure. The lattice volume of this alloy is smaller than earlier reported $Cu_2MnSb$ ($T_N \sim 38K$) and $Ni_2MnSb$ ($T_c \sim 340K$).

## Magnetic studies

Variable temperature field cooled (FC) magnetization from 300 - 1000 K in an applied magnetic field of 500Oe for MnNi$_{0.9}$Cu$_{1.1}$Sb, MnNiCuSb and MnNi$_{1.1}$Cu$_{0.9}$Sb is shown in figure 3. The plot clearly shows a ferromagnetic transition. It is observed that $T_C$ decreases with increase in Ni content. $T_C$ is observed to be 705 K, 690 K and 670 K in MnNi$_{0.9}$Cu$_{1.1}$Sb, MnNiCuSb and MnNi$_{1.1}$Cu$_{0.9}$Sb respectively. The M vs H plot (shown in figure 4) for all the three composition at 5 K shows a typical ferromagnet like hysteresis loop.

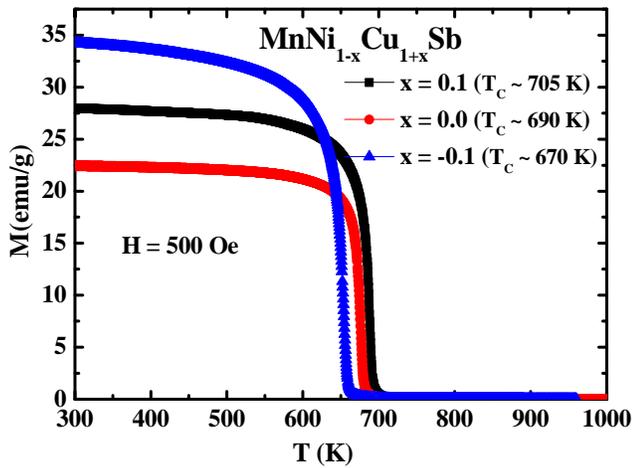

Figure3. Variable temperature magnetization plots for MnNi$_{1-x}$Cu$_{1+x}$Sb (x = 0.1, 0 and -0.1).

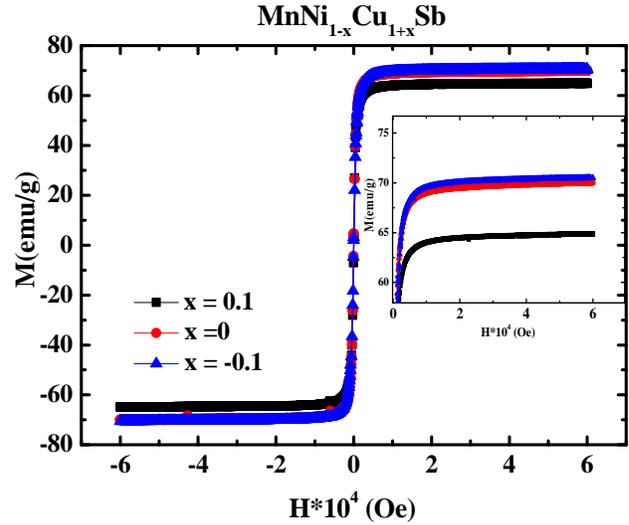

Figure4. Isothermal magnetization of MnNi$_{1-x}$Cu$_{1+x}$Sb (x = 0.1, 0 and -0.1) at 5K. Inset show the zoomed magnetization cycle.

Low value of coercive field in all the three composition shows them to be soft magnet. The magnetic moment increases with increase in Ni content and is 3.57$\mu_B$/f.u., 3.85$\mu_B$/f.u. and 3.87$\mu_B$/f.u. at 5K in MnNi$_{0.9}$Cu$_{1.1}$Sb, MnNiCuSb and MnNi$_{1.1}$Cu$_{0.9}$Sb respectively. This value of magnetic moment on Mn atom is close to earlier reports in Mn containing Heusler alloys. It has been found in earlier reports, by Neutron powder diffraction studies that in Mn containing Heusler alloy (Mn$_2$NiGa, Ni$_2$MnSb, Ni$_2$MnSn) [30, 31] magnetic moment resides only on Mn atoms and the other transition metal atom Ni does not carry any magnetic moment.

## Conclusion

We have synthesized an ordered quaternary 1:1:1:1 MnNiCuSb Heusler alloy. It is unique among all the Heusler alloys known so far; it is one of the few known *non-Fe* containing crystallographically ordered cubic quaternary Heusler alloy, T$_C$ ~ 690K. We believe the magnetic moment resides only on Mn atoms (3.85µ$_B$/Mn) which needs to be determined from neutron diffraction studies. We suspect it possesses a high spin polarization. This alloy with high T$_C$ and low coercive field (nearly square hysteresis loop) may be useful in applications in spintronics. Microscopic studies such as neutron diffraction, Mossbauer and ZF-NMR as a function of temperature would be useful to further characterize the magnetic state of the alloy.

## Acknowledgements

A.K.G acknowledges DST, India (SB/S1/PC-58/2012) for financial support. Z.H. and G.S.T. acknowledge UGC, India (F/2-6/98(SA-1)) and CSIR, India (9186(1158)/13EMR-I) respectively for fellowships.

† Visiting scientist at Solid State and Nanomaterials Research Laboratory, Department of Chemistry, IIT Delhi, India.

## References

(1) Heusler, F. *Verh. dtsch. phys. ges.* **1903**, 5, 219.

(2) Drews, J.; Eberz, U.; Schuster, H.-U. *Journal of the Less Common Metals* **1986**, 116, 271–278.

(3) Groot, R. A. De; Mueller, F. M.; Engen, P. G. Van; Buschow, K. H. J. *Phys. Rev. Lett.* **1983**, 50 (25), 2024−2027.

(4) Potter, H. H. *Proc. Phys. Soc.* **1929**, 41, 135.

(5) Pearson, E. *Z. Phys.* **1929**, 57, 115−133.

(6) Bradley, A. J.; Rodgers, J. W. *Proc. R. Soc. A,* **1934**, 144, 340−359.

(7) Kim, K. W.; Rhee, J. Y.; Kudryavtsev, Y. V.; Prokhorov, V. G.; Kim, J. M; Lee, Y. P *J. Korean Phys. Soc.* **2011**, 59, 3064−3068.

(8) Galanakis, I.; Dederichs, P. H. *Phys. Rev. B* **2002**, 66, 174429.

(9) Ishida, S; Fujii, S; Kashiwagi, S; Asano, S. *J. Phys. Soc. Jpn.* **1995**, 64, 2152−2157.

(10) Antonov, V. N.; Dürr, H. A.; Kucherenko, Y.; Bekenov, L. V.; Yaresko, A. N. *Phys. Rev. B - Condens. Matter Mater. Phys.* **2005**, 72 (5), 1–12.

(11) Balke, B.; Fecher, G. H.; Kandpal, H. C.; Felser, C.; Kobayashi, K.; Ikenaga, E.; Kim, J. J.; Ueda, S. *Phys. Rev. B - Condens. Matter Mater. Phys.* **2006**, 74 (10), 104405.

(12) Chadov, S.; Qi, X.; Kübler, J.; Fecher, G. H.; Felser, C.; Zhang, S. C. *Nat. Mater.* **2010**, 9 (7), 541–545.

(13) Lin, H.; Wray, L. A.; Xia, Y.; Xu, S.; Jia, S.; Cava, R. J.; Bansil, A.; Hasan, M. Z. *Nat. Mater.* **2010**, 9 (7), 546–549.

(14) Xiao, D.; Yao, Y.; Feng, W.; Wen, J.; Zhu, W.; Chen, X. Q.; Stocks, G. M.; Zhang, Z. *Phys. Rev. Lett.* **2010**, 105 (9), 25–28.

(15) Sagar, J.; Yu, C. N. T.; Lari, L.; Hirohata, A. *J. Phys. D. Appl. Phys.* **2014**, 47 (26), 265002.

(16) Tezuka, N.; Ikeda, N.; Mitsuhashi, F.; Sugimoto, S. *Appl. Phys. Lett.* **2009**, 94 (16), 162504.

(17) Wen, Z.; Sukegawa, H.; Kasai, S.; Inomata, K.; Mitani, S. *Phys. Rev. Appl.* **2014**, 2 (2), 1–2.

(18) Heusler, F; Starck, W; Haupt, E *Verh. dtsch. phys. ges.* **1903**, 5, 220.

(19) Halder, M.; Yusuf, S. M.; Nigam, A. K.; Keller, L. *Phys. Rev. B* L **2011**, 84, 094435.

(20) Alijani, V.; Ouardi, S.; Fecher, G. H.; Winterlik, J.; Naghavi, S. S.; Kozina, X.; Stryganyuk, G.; Felser, C. *Phys. Rev. B* **2011**, 84, 224416.

(21) Klaer, P.; Balke, B.; Alijani, V.; Winterlik, J.; Fecher, G. H.; Felser, C.; Elmers, H. J. *Phys. Rev. B* **2011**, 84, 144413.

(22) Nehra, J.; Sudheesh, V. D.; Lakshmi, N.; Venugopalan, K. *Phys. Status Solidi - Rapid Res. Lett.* **2013**, 7 (4), 289–292.

(23) Dai, X.; Liu, G.; Fecher, G. H.; Felser, C.; Li, Y.; Liu, H. *J. App. Phys.* **2014**, 901, 103–106.

(24) Halder, M.; Mukadam, M. D.; Suresh, K. G.; Yusuf, S. M. *J. Magn. Magn. Mater,* **2015**, 377, 220-225.

(25) Alijani, V.; Winterlik, J.; Fecher, G. H.; Naghavi, S. S.; Felser, C.*Phys. Rev. B* L **2011**, 83, 184428.

(26) Basit, L.; Fecher, G. H.; Chadov, S.; Balke, B.; Felser, C. *Eur. J. Inorg. Chem*, **2011**, 3950-3954.

(27) Özdoğan, K.; Şaşioğlu, E., Galanakis, I. *J. Appl. Phys.* **2013**, 113, 193903.

(28) *TOPAS*, version 4.2; Bruker AXS: Karlsruhe, Germany, **2009**.

(29) Webster, P. J. *J. Phys. Chem. Solids.* **1971**, 32, 1221.

(30) Brown, P. J.; Kanomata, T.; Neumann, K.; Neumann, K. U.; Ouladdiaf, B.; Sheikh, A; Ziebeck, K. R. A. *J. Phys. Condens. Matter* **2010**, 22 (50), 506001.

(31) Meinert, M; Schmalhorst, J. –M.; Reiss, G. *J. Phys. Condens. Matter* **2011**, 22, 116005.